\documentclass[prl,preprint,draft,amsmath,showpacs]{revtex4}
\usepackage{graphics}

\newcommand{\abs}[1]{\ensuremath{\left| #1 \right|}}
\newcommand{\nax}{Na$_x$CoO$_2$~}
\newcommand{\na}{Na$_{0.7}$CoO$_2$~}

\newcommand{\refl}{R($\omega$)~}
\newcommand{\sig}{$\sigma_1(\omega)$~}
\begin{document}
%%%%%%%%%%%%%%%%%%%%%%%%%%%%%%%%%%%%%%%%%%

\bibliographystyle{prsty}
\input epsf

\title {Optical evidence for the proximity to a spin-density-wave 
metallic state in \na}

\author {G. Caimi}
\author {L. Degiorgi}
\affiliation{Laboratorium f\"ur Festk\"orperphysik, ETH
Z\"urich,
CH-8093 Z\"urich, Switzerland}\

\author {H. Berger}
\author{N. Barisic}
\author{L. Forr\'o}
\affiliation{Institut de physique de la mati\`ere complexe (IPMC), 
EPF Lausanne, CH-1015 Lausanne, Switzerland}\
\author {F. Bussy}
\affiliation{Institute of Mineralogy and    Geochemistry, University 
of Lausanne, CH-1015 Lausanne, Switzerland}\

\date{\today}

\begin{abstract}
We present the optical properties of \na single crystals, measured 
over a broad spectral range as a function of temperature ($T$).
The capability to cover the energy range from the far-infrared up to 
the ultraviolet allows us to perform reliable Kramers-Kronig 
transformation, in order to obtain the absorption spectrum (i.e., the 
complex optical conductivity). To the complex optical conductivity we 
apply the generalized Drude model, extracting the frequency 
dependence of the scattering rate ($\Gamma$) and effective mass 
($m^*$) of the itinerant charge carriers. We find that 
$\Gamma(\omega)\sim \omega$ at low temperatures and for $\omega > T$. 
This suggests  that \na is at the verge of a spin-density-wave 
metallic phase.
\end{abstract}


\maketitle

The discovery of superconductivity at 5 K in hydrated  sodium 
cobaltate\cite{Takada} has attracted considerable attention. How 
water inclusion triggers superconductivity in \nax is not fully 
understood yet. The investigation of non-hydrated sample is therefore 
of relevance and a considerable research effort has been devoted to 
\nax specimens with $x$ ranging between 0.3 and 0.75. As $x$ 
increases from 0.3, the ground state goes from a paramagnetic metal 
to a charge-ordered insulator for $x=0.5$ to a Curie-Weiss metal 
around 0.7 and finally to a weak-moment magnetically ordered state 
for $x>0.75$ (Ref. \onlinecite{foo}). This latter phase is supposed 
to be equivalent to a so called spin-density-wave (SDW) metal. 
Several recent investigations\cite{sales,sugiyama,moto}, based on 
magnetic, thermal and transport properties as well as muon spin 
spectroscopy, indicate the formation of a SDW metallic state for 
$x=0.75$.
%\begin{figure} [!h]
%\begin{center}
%\resizebox{9.0 cm}{!}{\includegraphics{RS_v2.eps}}
%\caption{(Color online) Reflectivity (top panel) and  real part \sig 
%of the optical conductivity (bottom panel) of \na at selected 
%temperatures. The four high frequency  absorptions in \sig are 
%labeled (see text). Inset: \refl at 10 K between 0 and 0.2 eV, 
%emphasizing the linear behavior of \refl at low energies.}
%   \label{rs}  
%\end{center}
%\end{figure}
%<<<<<<<<<<<<<<<<<<<<<<<<FIGURE 1>>>>>>>>>>>>>>>>>>>>>>>>>
\begin{figure}[t]
    \begin{center}
     \leavevmode
     \epsfxsize=13cm \epsfbox {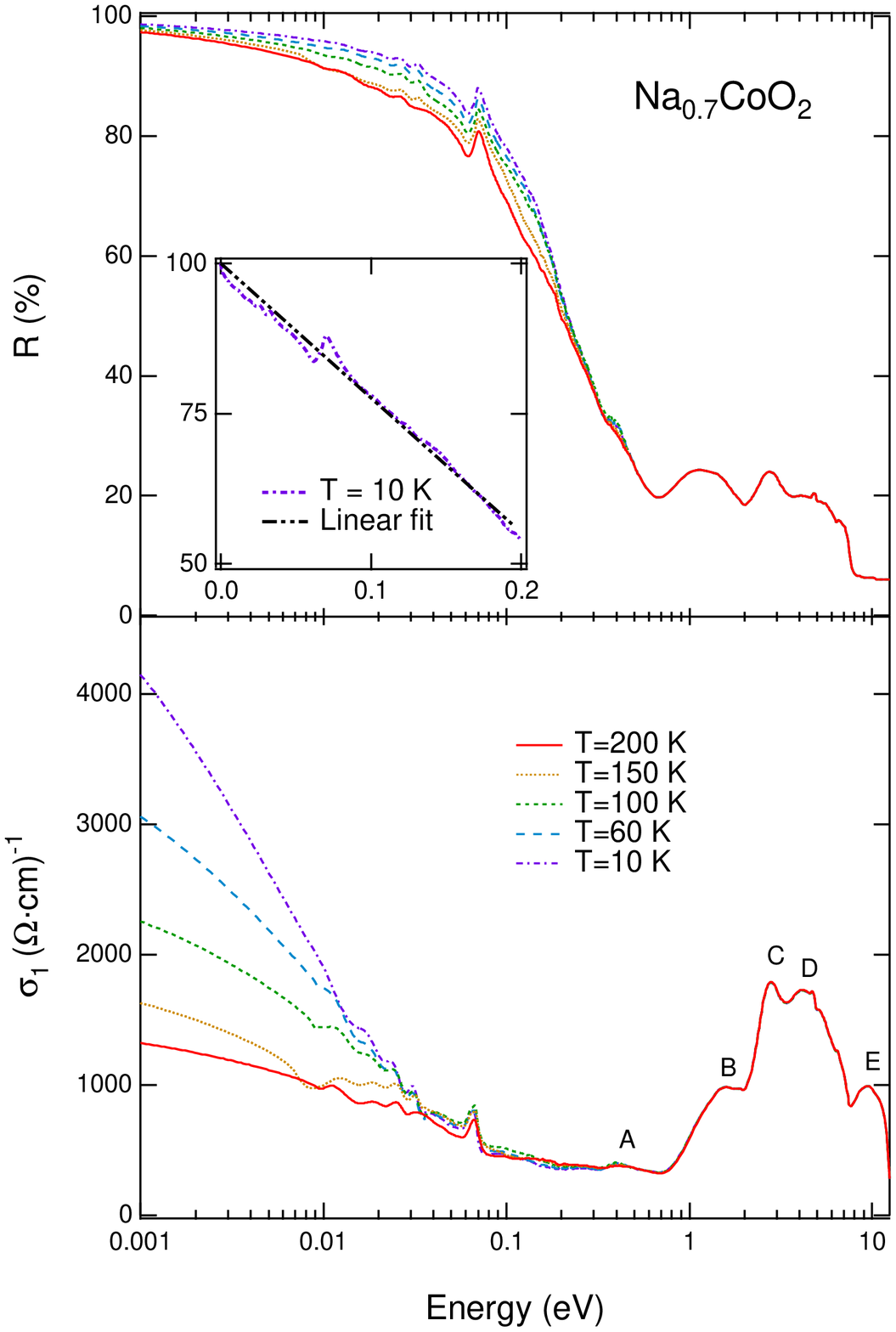}
      \caption{(Color online) Reflectivity (top panel) and  real part 
\sig of the optical conductivity (bottom panel) of \na at selected 
temperatures. The four high frequency  absorptions in \sig are 
labeled (see text). Inset: \refl at 10 K between 0 and 0.2 eV, 
emphasizing the linear behavior of \refl at low energies.}
\label{rs}
\end{center}
\end{figure}
%<<<<<<<<<<<<<<<<<<<<<<<<FIGURE 1>>>>>>>>>>>>>>>>>>>>>>>>>

Optical experiments are well-known tools in order to achieve 
information about the electrodynamic response of the investigated 
system, and to shed light on its electronic structure. From the 
absorption spectrum one can also extract the effective mass and the 
scattering rate of the itinerant charge carriers. Lupi $et~al.$ 
measured the optical conductivity of \nax (for $x=0.57$) and reported 
an ''anomalous Drude'' behavior, where the charge carrier have an 
effective mass of $5 m_b$  and their scattering rate is $\Gamma 
(\omega) = \frac{1}{\tau(\omega)}\sim \omega^{3/2}$ (Ref. 
\onlinecite{lupi}). Wang $et~al.$ optically investigated 
Na$_{0.7}$CoO$_2$, mainly focusing their attention on the high 
frequency spectral range. They found two broad interband transitions 
at 1.6 eV and 3.2 eV ascribed to the $t_{2g}-e_g$ band transitions by 
invoking the effect of the exchange  splitting\cite{wang}. 
Additionally, a mid-infrared peak at about 0.4 eV  was attributed to 
interband transition within the $t_{2g}$ manifold or to the 
electronic correlation effect. At lower frequencies a non simple 
Drude intraband response was detected, as well\cite{wang}.

In this short communication, we offer our optical investigations on 
high quality \na single crystals. Our specimens have a size 2 x 2 mm 
and were grown by the flux methods as thoroughly described 
elsewhere\cite{iliev}.  Samples from the same batch were furthermore 
characterized by the dc transport measurements\cite{forro}. The 
temperature dependence of the resistivity $\rho(T)$ (not shown here) 
within the $ab$ plane  displays a linear behavior in temperature from 
300 down to 100 K, as well as below 100 K, however  with a smaller 
slope, in agreement with data of Ref. \onlinecite{foo} and 
\onlinecite{ywang}. We performed optical reflectivity measurements 
from the far-infrared (FIR)  up to the ultraviolet (i.e., 5 meV- 12.4 
eV) as a function of temperature between 10 and 300 K. Measurements 
were performed also in a magnetic field up to 7 T. No changes in the 
spectra were however found as a function of field.  Our 
investigations covers the largest spectral range addressed so far on 
Na$_x$CoO$_2$. Details pertaining the experiments can be found 
elsewhere\cite{Wooten,Dressel}.

Figure \ref{rs} displays in the top panel the optical reflectivity 
R($\omega$). Beside some absorption at about and above 1 eV, one can 
recognize the quite sharp plasma edge feature with onset at $\simeq$ 
0.7 eV. \refl increases with decreasing temperature below 0.2 eV, 
indicative for the metallic character of Na$_{0.7}$CoO$_2$. 
Furthermore, we clearly see  at 0.07 eV the infrared-active phonon 
mode. Our data bears an overall similarity with the finding of Lupi 
$et~al.$ for Na$_{0.57}$CoO$_2$ (Ref. \onlinecite{lupi}). 
There is also a rough agreement with the optical response measured on 
\na by Wang  $et~al.$\cite{wang} In this latter data, there is a 
crossing of the \refl spectra around 0.1 eV and the formation of a 
bump at about 0.03 eV with decreasing temperature, for which neither  
our spectra nor those of  Lupi $et~al.$ give a clear cut evidence. 

The bottom panel of Fig. \ref{rs} shows the real part \sig of the 
optical conductivity.  We have extrapolated the \refl spectra towards 
zero energies  with the Hagen-Rubens 
extrapolation\cite{Wooten,Dressel}, using dc values of the 
conductivity in agreement with the transport results\cite{moto, foo, 
ywang}. Standard extrapolation of  $R(\omega)\sim \omega^{-s}$ (with 
$2<s<4$) were employed at high frequencies. It is easy verified that 
the main conclusions of our work  are fully independent from the 
employed extrapolations due to the extremely broad measured spectral 
range. As expected from the \refl spectra, the effective intraband 
metallic component in \sig is enhanced below 0.03 eV and the $ 
\sigma_1 (\omega\rightarrow0)$ limit increases with decreasing 
temperature, as typical for a metallic system.  
At higher frequencies we recognize a weak feature at 0.4 eV (A) 
followed by more pronounced and well defined features at 1.4 eV (B,) 
at 2.8 eV (C), at 4.8 eV (D), and at 10 eV (E). These latter 
absorptions are in fair agreement with the data of Wang  
$et~al.$\cite{wang} and in good accord with the recent findings 
obtained with the ellipsometry method\cite{bernhard}. Furthermore, 
angle resolved photoemission results identify electronic excitations 
at 0.7, 3, 4.1, 6 and 11 eV  (Ref. \onlinecite{Hasan}). In passing, 
we note that some of the detected absorptions could be ascribed to 
electronic interband transitions involving the Co  $3d$ $t_{2g}$ and 
$e_g$ manifolds and their exchange splitting\cite{singh,kunes}.

The metallic component of \sig cannot be fully reproduced by a simple 
Drude term, the most common description for simple metals, also 
successfully applied in several oxides. The optical conductivity can 
be alternatively described in terms of an "anomalous or generalized 
Drude" model, where both the effective mass $m^*(\omega)$ and the  
scattering rate $\Gamma(\omega)$ of the itinerant charge carriers are 
allowed  to depend on frequency. We analyze the complex optical 
conductivity $\tilde{\sigma} 
(\omega)=\sigma_1(\omega)+i\sigma_2(\omega)$ by applying the 
following expression:
\begin{equation}
\label{si}
\tilde{\sigma} 
(\omega)=\frac{\omega^2_p}{4\pi}\frac{1}{\Gamma(\omega)-i\omega\frac{m^*(\omega)}{m_b}}.
\end{equation}
By inverting  eq. (\ref{si}) we obtain:
\begin{eqnarray}
\Gamma(\omega) & = & 
\frac{\omega^2_p}{4\pi}\frac{\sigma_1}{\abs{\sigma}^2} \\
\frac{m^*(\omega)}{m_b} & = & 
\frac{\omega^2_p}{4\pi}\frac{\sigma_2}{\omega\abs{\sigma}^2}
\end{eqnarray}
In practice, we use $\tilde{\sigma}(\omega)$,  obtained by the 
Kramers-Kronig analysis, in order to evaluate the frequency 
dependence of the effective mass and scattering rate. $\omega^2_p$ is 
the spectral weight associated with the itinerant charge carriers, 
and it can be estimated by integrating \sig from zero frequency up to 
a cut-off frequency $\omega_c$ coinciding with onset of electronic 
interband transitions. We choose  $\omega_c\approx 0.63$ eV, giving a 
value of $\omega_p\simeq 1.17$ eV.
%\begin{figure} [!h]
%\begin{center}
%    \resizebox{9.0 cm}{!}{\includegraphics{Gamm_fit_paper_v2.eps}}
% \caption{(Color online) Frequency dependence of the scattering rate 
%and its fit according to eq. (\ref{pl}) at selected temperatures.  
%Note that the IR active phonon has been subtracted in order to better 
%highlight the linear or sub-linear fit.  Inset: the original curve of 
%$\Gamma(\omega)$ (i.e., comprehensive of the IR phonon at 0.07 eV) at 
%10 K is shown with the  fit after eq. (\ref{pl}) with $\alpha=1$. The 
%phonon subtraction does not affect the fit of $\Gamma(\omega)$.  This 
%is  true  at all temperatures. }
%\label{gamma}
%\end{center}
%\end{figure}
%<<<<<<<<<<<<<<<<<<<<<<<<FIGURE 2>>>>>>>>>>>>>>>>>>>>>>>>>
\begin{figure}[t]
    \begin{center}
     \leavevmode
     \epsfxsize=13cm \epsfbox {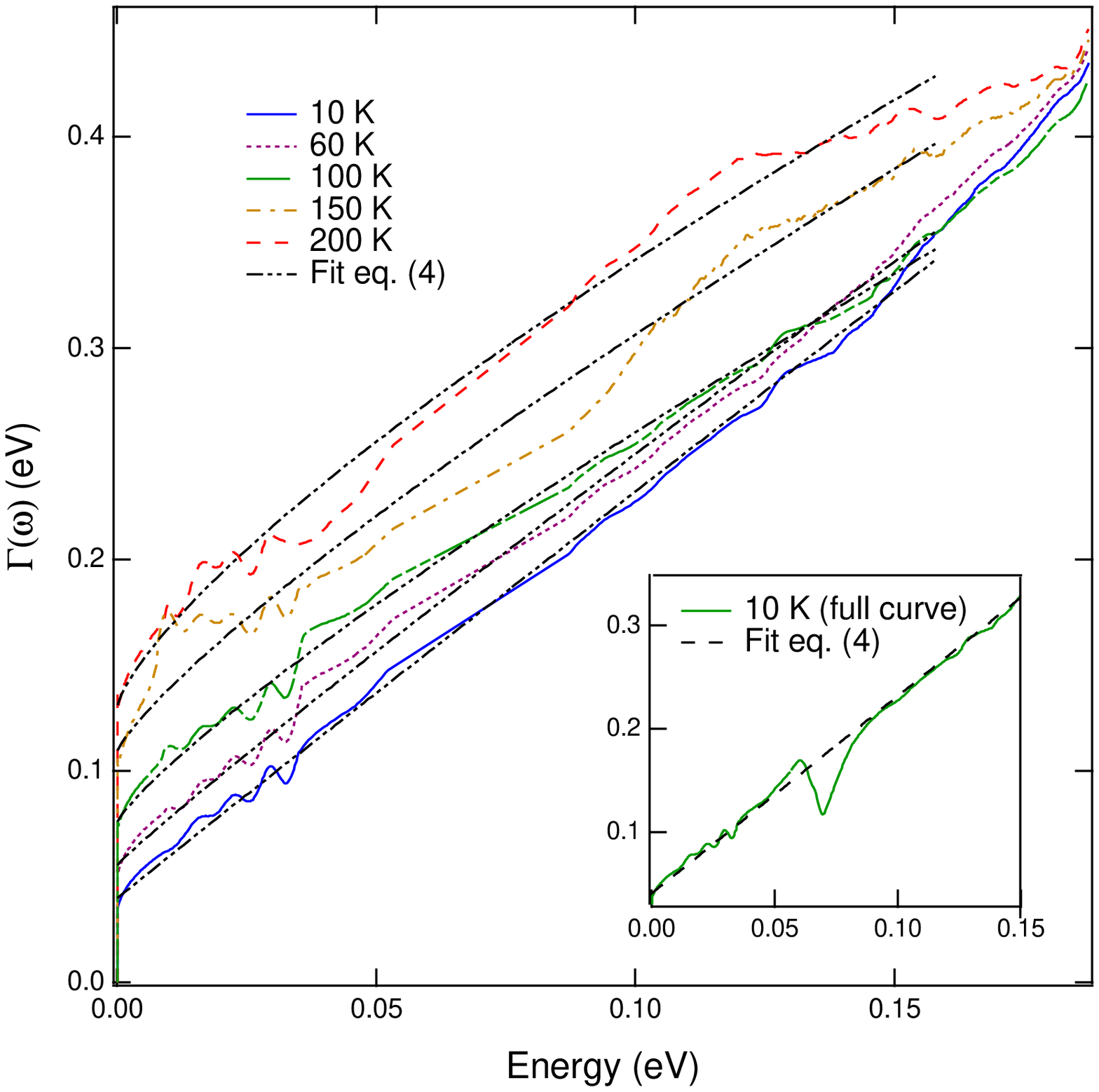}
      \caption{(Color online) Frequency dependence of the scattering 
rate and its fit according to eq. (\ref{pl}) at selected 
temperatures.  Note that the IR active phonon has been subtracted in 
order to better highlight the linear or sub-linear fit.  Inset: the 
original curve of $\Gamma(\omega)$ (i.e., comprehensive of the IR 
phonon at 0.07 eV) at 10 K is shown with the  fit after eq. 
(\ref{pl}) with $\alpha=1$. The phonon subtraction does not affect 
the fit of $\Gamma(\omega)$.  This is  true  at all temperatures.}
\label{gamma}
\end{center}
\end{figure}
%<<<<<<<<<<<<<<<<<<<<<<<<FIGURE 2>>>>>>>>>>>>>>>>>>>>>>>>>
Figure \ref{gamma} displays the frequency dependence of the 
scattering rate $\Gamma$ below 0.2 eV. For clarity, the depletion of 
$\Gamma(\omega)$ at 0.07 eV, due to the phonon mode, has been removed 
from the data (see below). We remark moreover  that $m^*(\omega)$ 
(not shown here) weakly increases with decreasing frequency at all 
temperatures, reaching a value of about $5 m_b$ in the limit 
$\omega\rightarrow0$. This agrees with Lupi's data\cite{lupi}.

\begin{figure}[t]
    \begin{center}
     \leavevmode
     \epsfxsize=11cm \epsfbox {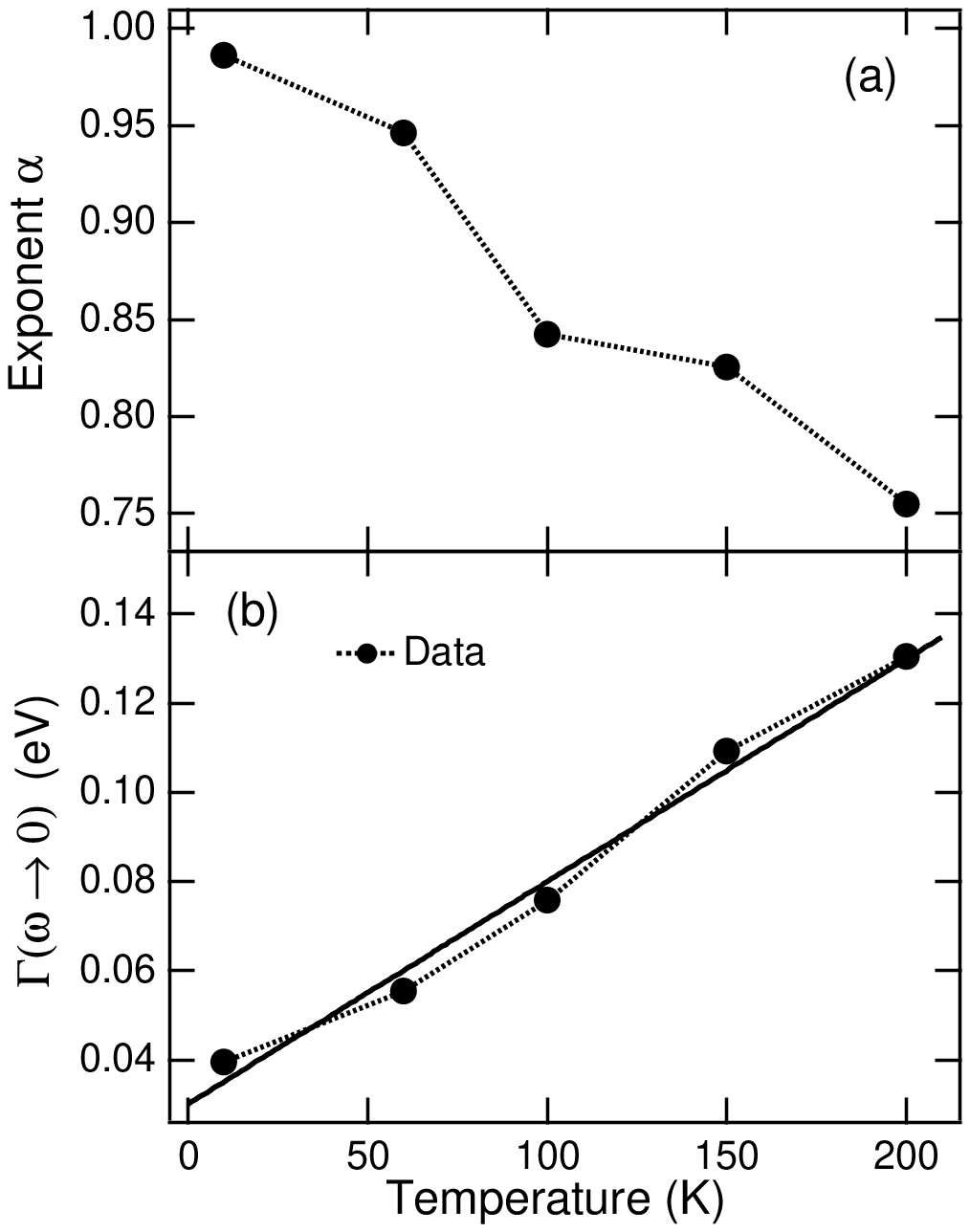}
      \caption{(a) Temperature dependence of the exponent $\alpha$  
in eq. (\ref{pl}). (b) Temperature dependence of $\Gamma (\omega)$ in 
the static limit $\omega\rightarrow 0$. $\Gamma(\omega\rightarrow 0)$ 
can be well approximated by a linear fit.}
\label{exp}
\end{center}
\end{figure}

As expected,  $\Gamma(\omega)$ decreases with decreasing temperature. 
Over a very broad spectral range, extending up to about 0.15 eV, 
$\Gamma(\omega)$ can be fitted with the power law expression:
\begin{equation}
\label{pl}
\Gamma(\omega)\sim \omega^{\alpha}.
\end{equation}
We establish that $\alpha\simeq1$ for  temperatures below about 50 K. 
$\Gamma(\omega)\sim \omega$ might be indicative of a non-Fermi liquid 
behavior in Na$_{0.7}$CoO$_2$.  The exponent $\alpha$ (Fig. 
\ref{exp}a) tends nevertheless to decrease  at higher temperatures 
(e.g., $\alpha=0.75$ at 200 K). We emphasize at this point that our 
fit after eq. (\ref{pl}) is not affected by the ad-hoc phonon 
subtraction.  The inset of Fig. \ref{gamma} shows indeed that even 
the original curve of $\Gamma(\omega)$ at 10 K can be well fitted by 
eq. (\ref{pl}) with $\alpha=1$. The same applies for all 
temperatures, making our analysis  of $\Gamma(\omega)$ robust.  The 
linear frequency dependence of $\Gamma(\omega)$ at $\omega> T $ pairs 
with the linear temperature dependence of $\rho_{dc} (T) $ for 
$T<100$ K in the compound with 71 \% Na content\cite{foo,forro}.

%\begin{figure} [!h]
%\begin{center}
%    \resizebox{7.0 cm}{!}{\includegraphics{Exp_Ggdc_v2.eps}}
%    \caption{(a) Temperature dependence of the exponent $\alpha$  in 
%eq. (\ref{pl}). (b) Temperature dependence of $\Gamma (\omega)$ in 
%the static limit $\omega\rightarrow 0$. $\Gamma(\omega\rightarrow 0)$ 
%can be well approximated by a linear fit.}
%     \label{exp}
%  \end{center}
%\end{figure}
%<<<<<<<<<<<<<<<<<<<<<<<<FIGURE 3>>>>>>>>>>>>>>>>>>>>>>>>>
%<<<<<<<<<<<<<<<<<<<<<<<<FIGURE 3>>>>>>>>>>>>>>>>>>>>>>>>>
Ruvalds and Virosztek proposed a while ago a Fermi-surface nesting 
scenario for describing the optical properties of superconducting 
oxides\cite{ruvald}. They showed that Fermi-surface nesting modifies 
the electron-electron scattering and therefore yields an unusual 
variation of the optical reflectivity. Within this scenario, also 
applicable for the charge- and spin-density-wave state where nesting 
is an essential ingredient\cite{ruvald}, the effective Drude 
component is characterized by a relaxation rate that is linear in 
frequency for  $\omega> T $ (Fig. \ref{gamma}).  Ruvalds and 
Virosztek\cite{ruvald} also predicts that $\Gamma(\omega\rightarrow 
0)\sim T $, as  verified in   Fig. \ref{exp}b, where  
$\Gamma(\omega)$ in the static limit $\omega\rightarrow 0$ as a 
function of $T$ is shown. Furthermore, \refl is linear in $\omega$ in 
a broad spectral range (inset of Fig. \ref{rs}), in agreement with 
the theory\cite{ruvald}.  Our findings in \na  support the 
Fermi-surface nesting scenario. Consequently, considering the phase 
diagram reported in Ref. \onlinecite{foo}, we can then affirm that 
\na seems  to be at the verge of a SDW metallic phase. The linear 
frequency dependence of $\Gamma(\omega)$ at low temperatures differs 
from the conclusion of Lupi $et~al.$ on Na$_{0.57}$CoO$_2$, for which 
$\alpha\simeq3/2$ (Ref. \onlinecite{lupi}). This could be explained 
by the different stoichiometry of the samples. Na$_{0.57}$CoO$_2$ is 
quite close to the charge-ordered insulating boundary (at $x=0.5$) 
between a paramagnetic metal and a Curie-Weiss  metal\cite{foo}. On 
the contrary, our sample  \na  is located in the phase 
diagram\cite{foo} well  within the Curie-Weiss metal sector and is 
close to the boundary (at least for low temperatures) of the SDW 
metallic phase. We shall finally note that after Ref. 
\onlinecite{ruvald} a non Fermi liquid-Fermi liquid crossover is not 
excluded at very low temperatures in the case of an imperfect 
Fermi-surface nesting. This would be  in agreement with the 
dc-transport data on Na$_{0.7}$CoO$_2$. Indeed for $T<1$ K,  
$\rho(T)$ displays a typical Fermi liquid $T^2$ behavior\cite{li}.

In conclusion, we have provided the complete absorption spectrum of 
Na$_{0.7}$CoO$_2$. The frequency dependence of the scattering rate of 
the itinerant charge carriers is extracted  from the complex optical 
conductivity, and we have established that $\Gamma(\omega)\sim 
\omega$ at low temperatures.  \na seems to be in the proximity of a 
spin-density-wave metallic state. It turns out\cite{foo} that the 
exact stoichiometry plays an essential role in defining the intrinsic 
physical properties of Na$_x$CoO$_2$. As future outlook, we wish to 
fine tuning the Na content  $x$ between 0.7 and 0.85 and study the 
corresponding optical response. The SDW metallic phase for 
$x\geq0.75$ is an appealing  scenario, supported by various 
experimental results, but might be not the unique one. Interestingly 
enough, Bernhard $et~al.$\cite{bernhard} found evidence for a 
polaronic band in the case of $x=0.82$.

\begin{acknowledgments}
The authors wish to thank J. M\"uller for technical help and  A.
Perucchi for fruitful discussions. This work
has been supported by the Swiss National Foundation for the
Scientific Research.\end{acknowledgments}


\begin{thebibliography}{10}

\bibitem{Takada}
K. Takada, H. Sakurai, E. Takayama-Muromachi, F. Izumi, R.~A. 
Dilanian, and T.
  Sasaki, Nature {\bf 422},  53  (2003).

\bibitem{foo}
M.~L. {Foo}, Y. {Wang}, S. {Watauchi}, H.~W. {Zandbergen}, T. {He}, 
R.~J.
  {Cava}, and N.~P. {Ong}, cond-mat/0312174.

\bibitem{sales}
B.~C. {Sales}, R. {Jin}, K.~A. {Affholter}, P. {Khalifah}, G.~M. 
{Veith}, and
  D. {Mandrus}, cond-mat/0402379.

\bibitem{sugiyama}
J. {Sugiyama}, J.~H. {Brewer}, E.~J. {Ansaldo}, H. {Itahara}, T. 
{Tani}, M.
  {Mikami}, Y. {Mori}, T. {Sasaki}, S. {Hebert}, and A. {Maignan},
  cond-mat/0310516.

\bibitem{moto}
T. Motohashi, R. Ueda, E. Naujalis, T. Tojo, I. Terasaki, T. Atake, M.
  Karppinen, and H. Yamauchi, Phys. Rev. B {\bf 67},  064406  (2003).

\bibitem{lupi}
S. {Lupi}, M. {Ortolani}, and P. {Calvani}, cond-mat/0312512.

\bibitem{wang}
N.~L. {Wang}, P. {Zheng}, D. {Wu}, Y.~C. {Ma}, T. {Xiang}, R.~Y. 
{Jin}, and D.
  {Mandrus}, cond-mat/0312630.

\bibitem{iliev}
M.~N. Iliev, A.~P. Litvinchuk, R.~L. Meng, Y. Sun, J. Cmaidalka, and 
C.~W. Chu,
  Physica C {\bf 402},  239  (2004).

\bibitem{forro}
The dc resistivity $\rho (T)$ was measured within the $ab$ plane and 
along the
  $c$-axis at EPF-Lausanne, with the conventional four points contact 
method.

\bibitem{ywang}
Y. Wang, N.~S. Rogado, R.~J. Cava, and N.~P. Ong, Nature {\bf 423},  
425
  (2003).

\bibitem{Wooten}
F. Wooten, {\em Optical Properties of Solids} (Academic Press, New 
York, 1972).

\bibitem{Dressel}
M. Dressel and G. Gr\"uner, {\em Electrodynamics of Solids} 
(Cambridge University
  Press, Cambridge, 2002).

\bibitem{bernhard}
C. {Bernhard}, A.~V. {Boris}, N.~N. {Kovaleva}, G. {Khaliullin}, A. 
{Pimenov},
  L. {Yu}, D.~P. {Chen}, C.~T. {Lin}, and B. {Keimer}, 
cond-mat/0403155.

\bibitem{Hasan}
M.~Z. {Hasan}, Y.~D. {Chuang}, A.~P. {Kuprin}, Y. {Kong}, D. {Qian}, 
Y.~W.
  {Li}, B.~L. {Mesler}, Z. {Hussain}, A.~V. {Fedorov}, R. 
{Kimmerling}, E.
  {Rotenberg}, K. {Rossnagel}, H. {Koh}, N.~S. {Rogado}, M.~L. {Foo}, 
and R.~J.
  {Cava}, cond-mat/0308438.

\bibitem{singh}
D.~J. Singh, Phys. Rev. B {\bf 61},  13397  (2000).

\bibitem{kunes}
J. {Kunes}, K.~W. {Lee}, and W.~E. {Pickett}, cond-mat/0308388  .

\bibitem{ruvald}
J. Ruvalds and A. Virosztek, Phys. Rev. B {\bf 43},  5498  (1991).

\bibitem{li}
S.~Y. {Li}, L. {Taillefer}, D.~G. {Hawthorn}, M.~A. {Tanatar}, J. 
{Paglione},
  M. {Sutherland}, R.~W. {Hill}, C.~H. {Wang}, and X.~H. {Chen},
  cond-mat/0401099.

\end{thebibliography}
\end{document}